\documentclass[12pt]{article}
\usepackage{amsmath,amssymb}
\usepackage{graphicx}
\usepackage{a4wide}
\usepackage[english]{babel}
\usepackage[colorlinks,linkcolor=blue,anchorcolor=blue,citecolor=blue,backref=page]{hyperref}
\usepackage{tikz}

\usetikzlibrary{shapes.geometric, arrows, calc}

\tikzstyle{process} = [rectangle, minimum width=3cm, minimum height=1cm, text centered, draw=black]
\tikzstyle{arrow} = [thick,->,>=stealth]

\def\leq{\leqslant}
\def\geq{\geqslant}
\def\({\left(}
\def\){\right)}

\usepackage[norefs,nocites]{refcheck}
\long\def\beginFORGET#1\endFORGET{}

\let\phi\varphi
\let\epsilon\varepsilon

\def\qed{{\hskip0pt\unskip\unskip\nobreak\hfil\penalty50
          \hskip1em\hbox{}\nobreak\hfil
           {$\square$}
          \parfillskip=0pt\finalhyphendemerits=0
          \par}\medskip}

               {\noindent{\bf Proof.}\ }
               {\qed}

               {\noindent{\bf Proof of #1.}\ }
               {\qed}

\newcommand{\BZ}{{\mathbb{Z}}}

\newcommand{\CU}{{\mathcal U}}

\newbox\mybox
\def\arrover#1{\mathrel{
       \setbox\mybox=\hbox spread 1.4em
              {\hfil$\scriptstyle#1$\hfil}
       \vbox{\offinterlineskip\copy\mybox
             \hbox to\wd\mybox{\rightarrowfill}}}}

\def\larrover#1{\mathrel{
       \setbox\mybox=\hbox spread 1.4em
              {\hfil$\scriptstyle#1\vphantom{g}$\hfil}
       \vbox{\offinterlineskip\copy\mybox
             \hbox to\wd\mybox{\leftarrowfill}}}}

\def\ontoover#1{\mathrel{
       \setbox\mybox=\hbox spread 1.4em
              {\hfil$\scriptstyle#1\vphantom{g}$\hfil}
       \vbox{\offinterlineskip\copy\mybox
             \hbox to\wd\mybox{\rightarrowfill\hskip-2.8mm
                               $\rightarrow$}}}}
\def\leftontoover#1{\mathrel{
       \setbox\mybox=\hbox spread 1.4em
              {\hfil$\scriptstyle#1\vphantom{g}$\hfil}
       \vbox{\offinterlineskip\copy\mybox
             \hbox to\wd\mybox{$\leftarrow$\hskip-2.8mm
                               \leftarrowfill}}}}

\def\Pet{\mathit{Pet}}

\begin{document}

\title{SeCritMass: Threshold Secret Petitions}

\author{Florian Breuer \\
School of Information and Physical Sciences, \\
University of Newcastle, \\
Newcastle, NSW 2308,  Australia \\
florian.breuer@newcastle.edu.au}

\maketitle

\begin{abstract}
	We introduce the notion of an $n$-threshold secret petition, in which users add encrypted signatures to a petition, and the signatures are decrypted if and only if at least $n$ signatures have been gathered. This solves the coordination problem in which users wish to sign a petition or commit to a cause, but do not want to be identified as having signed it before enough others have signed it too.
	
	We present an implementation of such a petition based on the ElGamal cryptosystem. 
	
	Applications include reporting misconduct in situations were complainants hesitate to come forward alone, such as in allegations of sexual harassment or police brutality.
\end{abstract}

%
%

\section{Introduction}

Imagine you work at MegaCorp and you and many of your colleagues are convinced that the CEO is running the company into the ground. How do you convince the board of directors to act? If you contact them, your lone voice is likely to be ignored and the CEO, who is known for her vindictiveness, will fire you. 

The board would be more inclined to take you seriously if you send them a petition signed by at least 100 employees. But how do you start such a petition? The CEO has informants among your colleagues, and you can be sure that if you start such a petition, the CEO will find out and fire you along with the first few colleagues who have signed the petition. As a result, nobody dares to sign your petition unless many others have signed it first - too many for the CEO to fire. This is a classic coordination problem.

The solution to this dilemma is to start a secret petition. The text of the petition is public, but the author and list of signatories is kept secret. An expiration date is set. If the petition gathers 100 signatures before the expiration date, the petition is published along with a full list of signatures. Now the petition has acquired a critical mass of signatures; the CEO cannot fire everybody and the board can no longer ignore it.

For this to work, the signatories need to be confident that the system (especially their anonymity if critical mass is not reached) is secure against attack by the CEO or other adversaries.

\section{Formalization}

For a positive integer $n$, we define an $n$-{\bf threshold secret petition} to be a petition which gathers encrypted signatures by valid users, which signatures can be decrypted once at least $n$ signatures have been gathered. 

Threshold secret petition schemes share some similarities with cryptographic voting systems, also known as e-voting systems. Many e-voting systems have been proposed, see \cite{Kho} for a survey. 

In an e-voting system, the users (voters) are first validated in a registration phase, then their votes are recorded in a way that each user can only vote once and their selection itself remains confidential. Finally, the number of votes for each option are tallied and announced. 

A threshold secret petition system is similar to an election where the users vote for or against the proposed petition, but the principal difference is that in an e-voting system the identities of those who voted in favour are never revealed, even if the petition gathers enough `yes' votes.

\subsection{Requirements}

The following requirements are desirable in an $n$-threshold secret petition scheme. 
\begin{description}
	\item[Pre-threshold anonymity.] Before $n$ signatures have been gathered, all signatures remain anonymous.
	\item[Post-threshold decryption.] If at least $n$ valid signatures have been gathered, then the full list of signatures can be decrypted by everybody. We say the petition has been triggered.
	\item[User validity.] Only eligible users with verified identities are able to sign, and once decrypted, their identity is not in doubt. No fake or impersonated signatures are accepted.
	\item[User uniqueness.] No user can sign the petition more than once.
	\item[Irrevokability.] Once added to the petition, no signature can be revoked.
	\item[Coercion resistance.] Before the petition has been triggered, an attacker cannot coerce a user to prove that they did or did not sign the petition. 
\end{description}


\subsection{SeCritMass scheme details}
In this section we describe a proof of concept ElGamal-based scheme to implement an $n$-threshold secret petition system.

The petition consists of a plaintext petition $\Pet$, along with an append-only chain of cyphersignatures. 
Each cyphersignature contains the encrypted identity of one signing user together with one fragment of the decryption key.
When $n$ key fragments have been published, all cyphersignatures can be decrypted.

\subsubsection{Entities}
The scheme we describe below involves the following entities.
\begin{description}
	\item[Author] composes the petition anonymously and may or may not be its first signatory.
	\item[Users] sign the petition, optionally adding personal testimony. Remain securely anonymous until a threshold number of users have signed the petition.
	\item[Validators] verify the identity and eligibility of users.
	\item[Key rabbits]\footnote{See Barry Hughart, {\em The Bridge of Birds}} hold shares of key fragments, and publish a key fragment along with a user's encrypted signature when authorized to do so by a validator.
\end{description}

\subsubsection{Parameters.} 
\begin{itemize}
	\item $n$ : The threshold number of signatures needed to trigger the petition. 
	\item $k$ : The number of key rabbits each holding one share of each key fragment.
	\item $t$ : The number of key rabbits required to recover one key fragment, $t \leq k$.
	\item $v$ : The number of validators who need to validate each user. 
\end{itemize}

\subsubsection{Key generation} 
Let $G$ be a cyclic group of order $q$ and generated by $g$ for which an efficient implementation exists and for which the discrete logarithm is assumed to be computationally unfeasible. 
Our first goal is to create an ElGamal \cite{ElG85} secret key $s\in \BZ_q := \BZ/q\BZ$ and public key $p=g^s\in G$. The secret key will be the sum of $n$ key fragments,
$s = s_1 + s_2 + \cdots + s_n$ in $\BZ/q\BZ$, with each fragment $s_i$ revealed alongside one encrypted user signature. 

For each $j=1,2,\ldots,n$, the fragment $s_j$ is created and distributed as follows.

\begin{enumerate}
\item Each key rabbit $K_i$ chooses a secret $x_{i,j} \in \BZ_q$ uniformly at random. 
\item Next, $K_i$ shares $x_{i,j}$ among the other key rabbits using a publicly verifiable secret sharing scheme with threshold $t$.
Importantly, as part of this process $K_i$ publishes $g^{x_{i,j}}$ and it is verified that this value indeed corresponds to the shared secret $x_{i,j}$.
Suitable secret sharing schemes are proposed in \cite{Pen12, Sta96}. 

Now, when a user wishes to sign the petition, any subset of $t$ key rabbits can collaborate to recover the secrets $x_{1,j}, x_{2,j}, \ldots, x_{k,j}$ and in particular compute the key fragment
$s_j = x_{1,j} + \cdots + x_{k,j}$.

\item To compute the public key, each key rabbit $K_i$ has published $g^{x_{i,1}}, \ldots, g^{x_{i,n}}\in G$ and the public key is computed as
\[
p = g^s = g^{\sum_{i,j} x_{i,j}} = \prod_{i,j}g^{x_{i,j}}.
\]

All of these computations are publicly verifiable by anyone, so there is no need to trust the integrity of any party at this step.


\item
The corresponding ElGamal \cite{ElG85} public encryption function is given by 
\[
E : G \rightarrow G\times G; \quad m \mapsto (g^y, mp^y),
\]
where $y\in\BZ_q$ is chosen uniformly at random for each message $m\in G$. Decryption is possible with the secret key $s$ by
\[
E^{-1} : G\times G \rightarrow G; \quad (g^y, mp^y) \mapsto m = mp^y\cdot (g^y)^{-s}.
\]

\item
Finally, the plaintext petition $\Pet$ is published alongside the above public key and petition parameters:
\[
(\Pet, G, g, p, n, k, t)
\]
\end{enumerate}

\subsubsection{Signatures.} It is crucial that the identity of each user who has signed the petition becomes known and undeniable after the petition has triggered. In the scheme described here, this is achieved by the use of validators, authorities who confirm the identity of each user, assigning to them a unique identity $u$. To defend against a dishonest validator certifying a fake user, we require at least $v \geq 1$ validators to validate each user. Also, the validators are not in possession of any key fragments and the validators do not know {\em which} petition the user intends to sign.

Each validator $V$ has a public hash function $h_V$.

To ensure each signing user is unique, we need a hash function $h_P$ associated to the petition, and publish $h_P(u)$ along with each cyphersignature. A user is only accepted if the hash $h_P(u)$ has not been published with any previous cyphersignature.

To ensure coercion resistance, however, the hash function $h_P$ needs to be computable only by a trusted authority. We propose distributing the computation of $h_P(u)$ among the key rabbits using a suitable form of secure multi-party computation \cite{Zha19}. Specifically, each key rabbit $K_i$ holds a share of the hash key for $h_P$ and receives a share $u_i$ of $u$ from the validator, and then a subset of $t$ key rabbits collaborates to compute $h_P(u)$. Hash functions suitable for efficient multi-party computation are discussed for example in \cite[Section~6]{BST20}.

%

\begin{enumerate}
	\item A user $U$ wishes to be validated by a validator $V$. $U$ proves their identity to $V$, who creates a unique identifier $u \in \BZ_q$ corresponding to that user. Importantly, the same user cannot obtain a different value of $u$ from a different validator. For example, $u$ might encode $U$'s social security number, a widely accepted public key or $U$'s membership number if the petition is open only to members of a particular organisation.
	
	\item The user writes a message $m_U$, which may consist of personal testimony, encrypted via $E$, or just a string of random bits, and creates the concatenation $p_U = \Pet \| m_U$. The user then computes the hash~$h_V(p_U)$ and sends it to $V$.
	
	\item Using a suitable threshold secret sharing protocol, $V$ shares the secret $u$ among the key rabbits $K_i$, securely sending each $K_i$ the pair $(u_i, h_V(p_U))$, where $u_i$ is a share of~$u$.
	
	\item Using a secure channel, the user sends each $K_i$ the hash preimage $p_U$, thus demonstrating that they are the validated user corresponding to the secret $u$ now shared amongst the key rabbits. This also ensures that every $K_i$ knows the correct petition which is to be signed.
	
	\item Using a suitable secure multi-party computation protocol, the key rabbits jointly compute the distributed hash $h_P(u)$ from the inputs~$u_i$.

	
%
%
%
	
	\item The key rabbits publish $\big(E(u_1), E(u_2), \ldots, E(u_k), h_P(u)\big)$.
	
	\item If this has been repeated for at least $v$ different validators, and the hash $h_P(u)$ has not appeared more than $v$ times, then a subset of at least $t$ key rabbits jointly reconstruct an unpublished key fragment $s_j$ and append the following cyphersignature to the petition's signature chain:
	\[
	\big(E(m_U), E(u_1), E(u_2), \ldots, E(u_k), h_P(u), s_j\big)
	\]
	
	\item Optionally, the same user can repeat this process at a later stage with new testimony $m_U'$, in which case $E(m_U')$ is appended to $E(m_U)$ in the above cyphersignature. However, this functionality may leave the user vulnerable to coercion.
\end{enumerate}

\begin{figure}\label{fig1}
	\begin{tikzpicture}[node distance=2cm]
		
		\node (signature_chain) [process] {Signature chain};
		\node (K1) [process, below left of=signature_chain, xshift=-2cm, yshift=-1cm] {$K_1$};
		\node (K2) [process, below of=signature_chain, yshift=0cm] {$K_2$};
		\node (K3) [process, below right of=signature_chain, xshift=2cm, yshift=-1cm] {$K_3$};
		\node (user) [process, below left of=K2, xshift=-2cm, yshift=-1cm] {User};
		\node (validator) [process, below right of=K2, xshift=2cm, yshift=-1cm] {Validator};
		
		\draw [arrow] (K1) -- (signature_chain);
		\draw [arrow] (K2) -- (signature_chain);
		\draw [arrow] (K3) -- (signature_chain);
		\draw [arrow] (user) -- (K1);
		\draw [arrow] (user) -- (K2);
		\draw [arrow] (user) -- (K3);
		\draw [arrow] (user) -- node[anchor=south] {} (validator);
		\draw [arrow] (validator) -- (K1);
		\draw [arrow] (validator) -- (K2);
		\draw [arrow] (validator) -- (K3);
		
		\node at ($(signature_chain)!0.5!(K1)$) [above] {};
		\node at ($(signature_chain)!0.5!(K2)$) [right] {};
		\node at ($(signature_chain)!0.5!(K3)$) [above] {};
		\node at ($(K2)!0.5!(user)$) [right] {};
		\node at ($(K1)!0.5!(user)$) [right] {};
		\node at ($(K3)!0.5!(user)$) [right] {};
		\node at ($(user)!0.5!(validator)$) [below] {$h_V(p_U)$};
		\node at ($(K1)!0.5!(validator)$) [right] {};
		\node at ($(K2)!0.5!(validator)$) [right] {};
		\node at ($(K3)!0.5!(validator)$) [right] {};
		\node at ($(user)!0.5!(K1)$) [left] {$p_U$};
		
		\node at ($(signature_chain)!0.5!(K3)$) [right=0.2cm] {$(E(m_U), E(u_1), E(u_2), E(u_3), h_P(u), s_j)$};
		\node at ($(K3)!0.5!(validator)$) [right=0.2cm] {$(u_i, h_V(p_U))$};
		
	\end{tikzpicture}
	\caption{Signature process with $k=3$ key rabbits and $v=1$ validator.}
\end{figure}
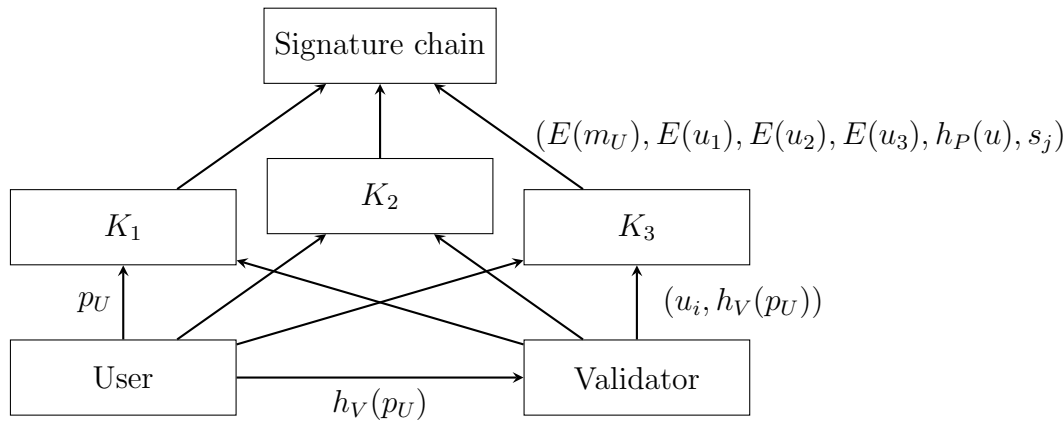

\subsubsection{Trigger.} Once $n$ cyphersignatures have been published, anybody can recover the secret key $s$ from the $n$ published key fragments and decrypt all cyphersignatures. In particular, the validated identity $u$ of the signing user can be reconstructed from the decrypted shares $u_1, u_2, \ldots, u_k$. The testimony $m_U$ can also be decrypted.

\subsubsection{Expiration.} The author may optionally set an expiration date, beyond which no further signatures are accepted and the petition cannot be triggered. This is achieved by requiring the key rabbits to delete their remaining secrets when the deadline arrives.

\subsubsection{Validators vs. key rabbits}
Why do we assign user validation and key fragment release to separate entities? The system would be simpler if the roles of validating a user and releasing a corresponding key fragment are performed by the same entity, in which case $v = t$. This may indeed be preferable in some situations. We treat the roles separately here because this allows choosing the parameters $v$ and $t$ separately, and also the knowledge of the user's identity and which petition is being signed is not held by the same party. Increasing $t$ decreases the likelyhood of key theft by a collection of $t$ corrupt key rabbits, but a large value of $v$ is more onerous on the users, and increases the risk that a user unwittingly approaches a corrupt validator and has their identity leaked.

\subsection{Applications and variants}

\subsubsection{Multiple thresholds}
Different users may be comfortable with different values of the threshold $n$. For example, Alice might want their signature to appear alongside $99$ or more other signatures ($n=100$), while Bob might be happy if the total threshold is at least $n=10$. To cater for such users, the author might set up a number of petitions, with identical wording, but with different secret keys and values of the threshold parameter $n$. Then each user can sign all of the variants with which they are comfortable. So in our example the author might set up variants with $n_1=10$, $n_2=50$ and $n_3=100$. Bob would sign all three variants, whereas Alice would sign only the $n=100$ variant.

Formally, we can model this as a single petition with a sequence of thresholds $n_1 < n_2 < \cdots < n_r$. Each user $U_i$ chooses a threshold $n_{j_i}$, the minimal threshold that user accepts. 
As soon as a subset $\CU$ of users has signed for which 
\[
\forall U_i \in \CU, \quad n_{j_i} \leq \#\CU,
\]
the petition will trigger along with the decrypted signatures of the users in $\CU$. 
If additional users sign on later, then it can trigger again for each larger set of users satisfying the above condition.

The above implementation may be modified as follows to implement multiple thresholds:
Each user $U$ chooses a threshold $n_U \in \{1, 2, \ldots, n\}$ and replaces the encryption function $E$ with $E_{n_U}$ corresponding to the public key
\[
p_{n_U} = \prod_{j=1}^{n_U} g^{s_j} = \prod_{i=1}^{k}\prod_{j=1}^{n_U}g^{x_{i,j}}.
\]
The corresponding cypher signature is then 
\[(E_{n_U}(m_U), E_{n_U}(u_1), E_{n_U}(u_2), \ldots, E_{n_U}(u_e), h_P(u), n_U, j, s_j),
\]
where $j$ is the largest index $j \leq n_U$ for which $s_j$ has not yet been published. 

This cyphersignature is decrypted with the secret key $s_1+s_2+\cdots +s_{n_U}$, which is available in this scheme if and only if the key fragments $s_1, s_2, \ldots, s_{n_U}$ have been published, i.e. at least $n_U$ users with thresholds $\leq n_U$ have signed.

\subsubsection{Sexual harassment reporting}

Suppose that a movie director has a nasty habit of auditioning new starlets on his ``casting couch''. Any starlet who complains will not be taken seriously as a lone voice, and loses her career. Instead, she can sign a threshold secret petition as above, encrypting her testimony along with her signature, and her cyphersignature is sent directly to the authorities. If enough starlets sign the petition, the authorities can decrypt the petition along with the testimonies and verified identities of all the signatories.

It is plausible that victims will be more willing to testify if they know that their testimony will only be decrypted if enough other victims come forward, too. This might also apply to other situations, such as allegations of police brutality.

In these examples, the petition need not be fully public, but rather the cyphersignatures can be sent directly to a relevant authority, such as an ombudsman, HR department or public prosecutor.

\subsubsection{Complaints within an organization}

In a large organization, an ombudsman might set up a complaints reporting system (e.g. for workplace sexual harassment) along these lines, effectively setting up one petition per staff member. Anybody wishing to make an encrypted complaint against a staff member can sign the relevant petition in full knowledge that their testimony will only be decrypted if enough other complaints are received against the same staff member. In such organisations it can be easier to validate users using their corporate identities, the threshold might be set low (e.g. $n=3$) and few validators and key rabbits are needed. In this situation, it is highly desirable that the validators do not know which petition is being signed.


%
%
%

\section{Attacks and limitations}

An $n$-threshold secret petition needs to guard against the following failure modes.
\begin{description}
	\item[Honeypot.] A corrupt author sets up a fake petition (e.g. distributing fake key fragments while keeping the secret key) to entrap users.
	
	\item[Key theft.] One or more key fragments are stolen (e.g. by compromising at least $t$ key rabbits), allowing an attacker to decrypt a petition before it triggers.
	
	\item[Invalid users.] One or more users are fake or impersonated (e.g. failure of at least $v$ validators). This can cause the petition to trigger before the required number of genuine signatures is gathered, violating pre-theshold anonymity.
	
	\item[Duplicate signatures.] One or more signatures is duplicated. As above, but the offending duplicate users are identified and may suffer reputational damage  when the petition is triggered.
	
	\item[User identity revealed.] If a validator is compromised, all users validated by the validator may be identified. 
	
	\item[User coercion.] An attacker coerces a user to prove that they have or have not signed, e.g. by gaining access to the hash function $h_P$.
	
	\item[Sabotage.] A petition cannot be triggered because key fragments have been lost, e.g. when more than $k-t$ key rabbits are compromised.
		
\end{description}
User confidence in the security of a petition system is critical and in particular users would be wary of a honeypot situation. For this reason, the key generation must use a publicly verifiable secret sharing scheme in which any party (e.g. a user) can verify both that the correct key fragments have been distributed and that nobody is in a position to know the secret key. This secures the system provided at most $t-1$ key rabbits are dishonest.


The integrity of the validators is clearly also crucial. A compromised validator can reveal the identities of all users they have validated, although they will not know which petitions were signed. Increasing the number of validators decreases the risk that all users are revealed in a single attack, but increases the number of targets of attack.

Perhaps the biggest danger in such a system is when many key rabbits (or validators) are controlled by a corrupt author. For this reason, it would be helpful if the key rabbits and validators belong to a dedicated petition organisation that has accumulated public trust over time.

A weakness of our implementation is coercion resistance: the user can prove that they have not signed the petition by going through the whole signing process and then showing that a new cyphersignature has been added to the signature chain. However, the user cannot prove that the new cyphersignature is indeed their own, since the encryption is performed by the key rabbits, and two cyphertexts $E(m_U)$ and $E'(m_U)$ of the same message are not equal in the ElGamal system. Partial protection against such coercion may thus be afforded if the key rabbits wait a random period of time before publishing a cyphersignature.

 Key to this is the fact that user duplication is avoided using a distributed hash function $h_P$; at least $t$ key rabbits need to collaborate to compute $h_P(u)$, so this cannot be computed by an attacker controlling fewer than $t$ key rabbits.

\subsubsection*{Acknowledgements.} The author would like to thank Erica Breuer and Hannes Breuer for productive discussions.


\end{document}